\documentclass[twocolumn,superscriptaddress]{revtex4-1}
\usepackage{amsmath,amssymb,bm,graphicx,hyperref,color}
\allowdisplaybreaks[1]

\def\gesim{\ \raise.3ex\hbox{$>$}\kern-0.8em\lower.7ex\hbox{$\sim$}\ }
\def\gesim{\ \raise.3ex\hbox{$>$}\kern-0.8em\lower.7ex\hbox{$\sim$}\ }

\newcommand\+{\dagger}

\newcommand\p{\bm{p}}
\newcommand\q{\bm{q}}
\newcommand\up{\uparrow}
\newcommand\down{\downarrow}

\newcommand\<{\langle}
\renewcommand\>{\rangle}

\begin{document}
\title{Mesoscopic spin transport between strongly interacting Fermi gases}

\author{Yuta Sekino$^*$}
\affiliation{Quantum Hadron Physics Laboratory, RIKEN Nishina Center, Wako, Saitama 351-0198, Japan}
\author{Hiroyuki Tajima$^*$}
\affiliation{Department of Physics, Kochi University, Kochi, Kochi 780-8520, Japan}
\affiliation{Quantum Hadron Physics Laboratory, RIKEN Nishina Center, Wako, Saitama 351-0198, Japan}
\author{Shun Uchino}
\affiliation{Waseda Institute for Advanced Study, Waseda University, Shinjuku, Tokyo 169-8050, Japan}
\thanks{These two authors contributed equally to this work.}
\date{\today}

\begin{abstract}
We investigate a mesoscopic spin current for strongly interacting Fermi gases through a quantum point contact.
Under the situation where spin polarizations in left and right reservoirs are same in magnitude but opposite in sign, we calculate the contribution of quasiparticles to the current by means of the linear response theory and 
many-body $T$-matrix approximation.
For a small spin-bias regime, the current in the vicinity of the superfluid transition temperature is strongly suppressed due to the formation of pseudogaps.
For a large spin-bias regime where the gases become highly polarized,
on the other hand, the current is affected by the enhancement of a minority density of states due to Fermi polarons.
We also discuss the broadening of a quasiparticle peak associated with an attractive polaron at a large momentum, which is relevant to the enhancement.
\end{abstract}

\maketitle

\section{\label{sec:introduction}Introduction}

Quantum simulation with ultracold atomic gases allows one to explore regimes of quantum-many body problems where
conventional systems such as condensed matter and nuclear matter are hard to reach~\cite{Bloch:2008,Giorgini:2008,Georgescu:2014}.  
A strongly-interacting Fermi gas realized with the Feshbach resonance is the prototype example, and
revealed existences of Bardeen-Cooper-Schrieffer (BCS) to Bose-Einstein condensation (BEC) crossover and  the unitarity regime in which
the typical length scale characterizing the atomic interaction disappears~\cite{Zwerger:2012,Strinati:2018,Ohashi:2020}.
Whilst both theoretical and experimental progresses have deepened understanding of the bulk phase structure,
understanding of the non-equilibrium properties remains challenging.

Recently, quantum transport of strongly-interacting Fermi gases attracts rising attention 
in association with the atomtronics devices where non-equilibrium properties on circuit or two-terminal systems are investigated~\cite{Krinner:2017}.
By using the controllable ultracold Fermi gases, quantum point contact~\cite{krinner:2015}, and junction systems attached with two~\cite{luick:2019} or three~\cite{valtolina:2015} dimensional reservoirs have experimentally been implemented.
In the case of the strongly-interacting superfluid regime, superfluid transport such as nonlinear current-bias characteristics induced by the multiple Andreev reflections~\cite{Husmann:2015}, and the AC Josephson oscillation~\cite{valtolina:2015} and 
DC Josephson effect~\cite{kwon:2019} has been confirmed.
In the case of the normal Fermi gas with a strong attractive interaction, a conductance beyond the quantized value has been found~\cite{Krinner:2016}.

In addition to controllability of the interaction,
transport systems with ultracold Fermi gases have an advantage that  spin transport can directly be measured with a tunable spin bias~\cite{Krinner:2016}. We note that this is in contrast to condensed matter systems where spin transport is normally measured in an indirect manner~\cite{Shen:2008}.  In the context of the interacting Fermi gases,
while mass transport contains information both on single-particle and collective excitations,
spin transport is not involved in Cooper pair transport~\cite{Uchino:2017,Liu:2017}.
Thus, it is expected that the spin transport measurement becomes a sensitive probe of single-particle excitations.

\begin{figure}[t]
\vspace{1cm}
\centering
\includegraphics[width=0.80\columnwidth,clip]{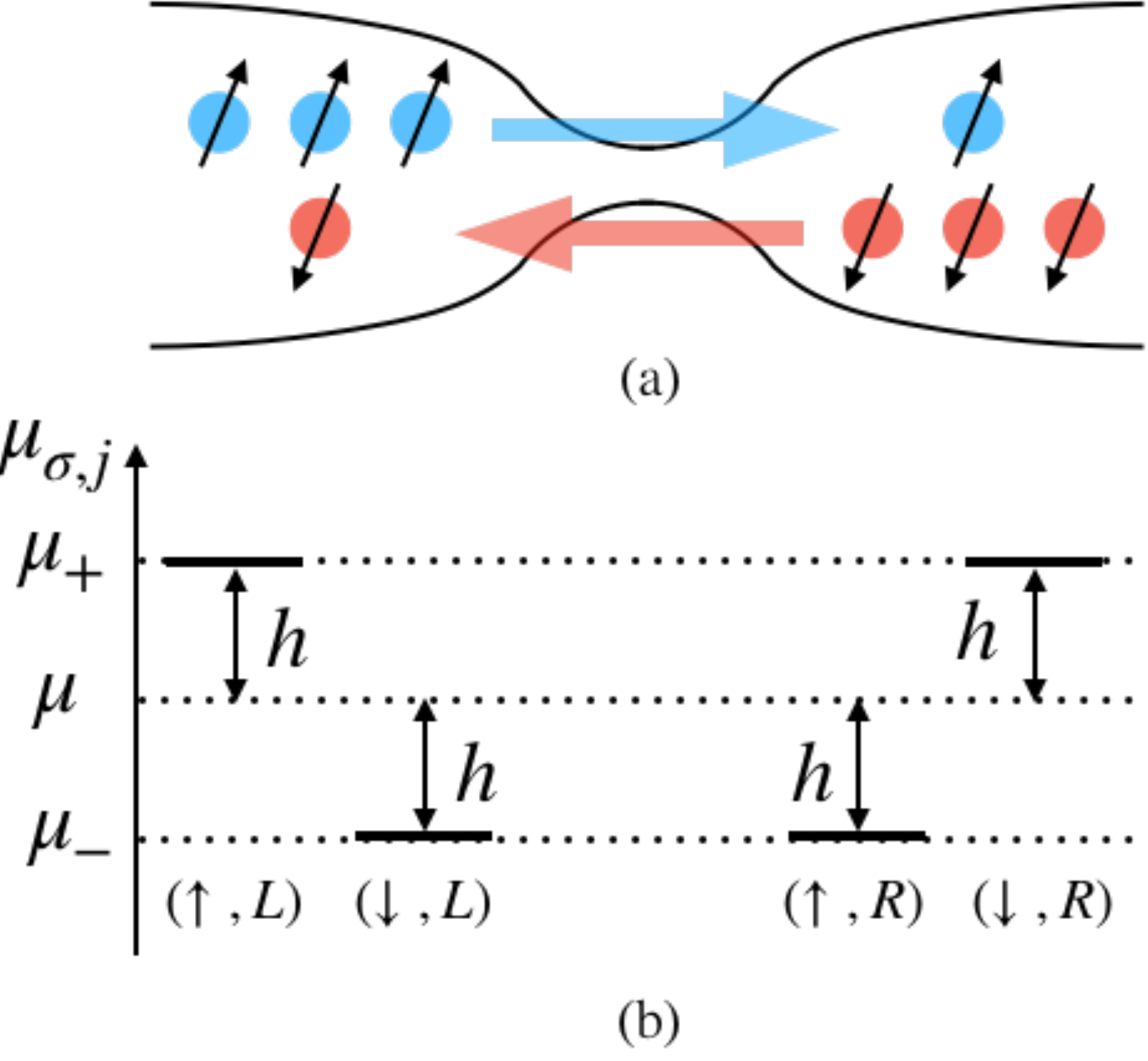}
\caption{\label{fig:setup}
Two terminal systems considered in this paper.
(a) Schematic view of the system.
Spin-up (spin-down) atoms are transported from left (right) to right (left).
(b) Chemical potentials for atoms in the left ($L$) and right ($R$) reservoirs.}
\end{figure}
In this paper, we investigate mesoscopic spin transport in attractively interacting two-component Fermi gases.
We consider the tunneling regime of
a two terminal system consisting of two normal Fermi gases through a quantum point contact (see Fig.~\ref{fig:setup}).
Spin polarizations of the Fermi gases in left and right reservoirs are assumed to be equal in magnitude but opposite in sign and are controlled by the parameter $h$.
Due to this spin imbalance, spin-up and spin-down fermions can move in opposite directions, 
and therefore a nonzero spin current without flow of a mass current is generated.

\begin{figure}[t]
\centering
\includegraphics[width=0.80\columnwidth,clip]{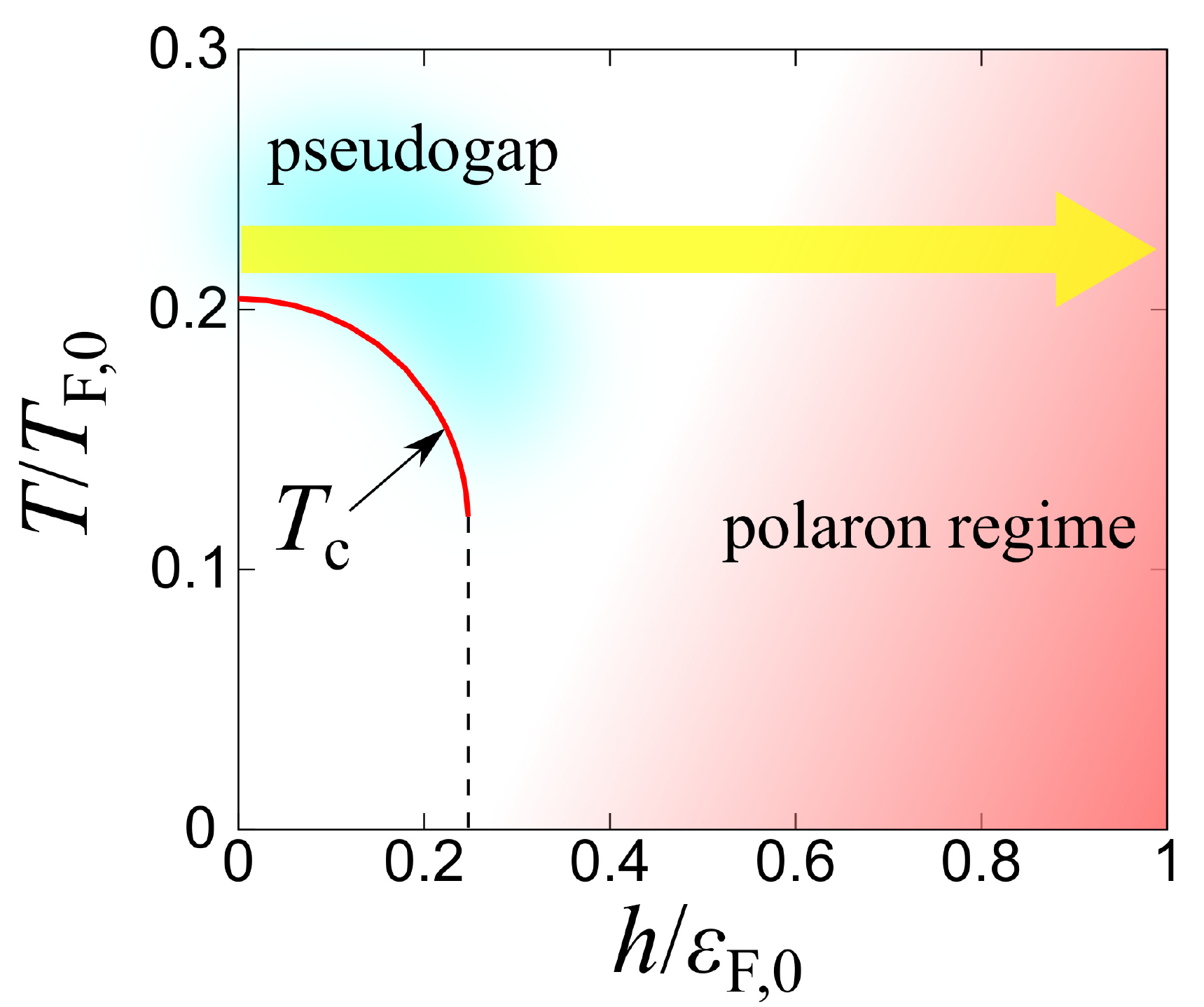}
\caption{\label{fig:phase}Phase diagram at unitarity in the plane of the temperature $T$ and the fictitious Zeeman field $h$ (see the main text in Sec.~\ref{sec:formalism} for details).
The superfluid transition temperature $T_\mathrm{c}$ (red solid line) is obtained in the ETMA .
In the dotted line, the second-order superfluid phase transition disappears~\cite{Kashimura:2012,Kashimura:2014}.}
\end{figure}
We focus on the strongly interacting regime near unitarity, where the absolute value of a $s$-wave scattering length $|a|$ is much larger than
the interatomic distance,
 and intercomponent fermions strongly interact with each other.
Figure~\ref{fig:phase} is the phase diagram of the spin-imbalanced Fermi gases at unitarity~\cite{Kashimura:2012,Kashimura:2014}.
When the low temperature and small $h$ regime is concerned, each reservoir becomes the superfluid where 
spin excitations are expected to be frozen.
Therefore, in this work, we elucidate spin transport above the superfluid critical temperature.
\par
One of the most remarkable features resulting from a strong interaction is the formation of a pseudogap~\cite{Mueller:2017,Jensen:2019,Tsuchiya:2009,Tsuchiya:2010,Watanabe:2010,Tsuchiya:2011,Mueller:2011,Magierski:2011,Su:2010}, where a dip structure in the single-particle density of states (DOS) in the normal phase near the transition temperature appears.
While this pseudogap phenomenon has indirectly been  observed  with photoemission spectroscopy~\cite{Stewart:2008,Gaebler:2010,Sagi:2015}, the measurement of thermodynamic quantities has showed a Fermi-liquid like behavior and suggested the absence of the pseudogap near unitarity~\cite{Nascimbene:2010,Nascimbene2:2010,Nascimbene:2011}.
For further understanding of  this system,
several quantities sensitive to the pseudogap as well as pairing fluctuations have been investigated~\cite{Kashimura:2012,Palestini:2012,Mink:2012,Enss:2012,Wlazlowski:2013,Pantel:2014,Tajima:2014,Tajima:2016,Tajima:2017,Jensen:2018,Tajima:2020},
including spin susceptibility measurements~\cite{Sommer:2011,Sanner:2011,Meineke:2012,Valtolina:2017}.
By looking at a small $h$ regime, we show that  the pseudogap effect can be captured with spin transport,
which may be complementary to  the tunneling spectroscopy in high-$T_{\rm c}$ superconductors~\cite{Fischer:2007}.
\par
In contrast, excitations of a highly polarized Fermi gas realized in the large $h$ regime is governed by Fermi polarons, which are mobile impurities immersed in a Fermi sea~\cite{Massignan:2013}.
Polaronic properties such as renormalization factors, effective masses, and polaron energies have been measured with RF spectroscopy~\cite{Schirotzek:2009,Nascimbene:2009,Kohstall:2011,Koschorreck:2012,Cetina:2016,Scazza:2017,Yan:2019,Ness:2020}.  Correspondingly, a bunch of theoretical works have been made~\cite{Chevy:2006,Combescot:2007,Combescot:2008,Pilati:2008,Prokofev1:2008,Prokofev2:2008,Vlietinck:2013,Kroiss:2015,Goulko:2016,Parish:2016,Punk:2009,Schmidt:2011,Kamikado:2017,Hu:2018,Tajima:2018,Tajima:2019,Mulkerin:2019,Mistakidis:2019,Liu:2019,VanHoucke:2020}, most of which
consider the zero-temperature and single-polaron limit by assuming that the impurity chemical potential is equal to the polaron energy. However, we note that the theory based on such an ideal limit is not available to 
 discuss polaronic effects in spin transport  at finite temperature and given chemical potential.
To incorporate the polaronic properties in a correct fashion, we  perform the finite-temperature many-body formalism of Fermi polarons.
By using this formalism, moreover,
we show that the crossover from the pseudogap to the polarons can be explored through spin transport.
\par
This paper is organized as follows.
In Sec.~\ref{sec:formalism}, we present the formalism of the tunneling Hamiltonian approach together with the diagrammatic $T$-matrix approximation.
Section~\ref{sec:result} is devoted to discussing how excitation properties in strongly interacting Fermi gases in reservoirs affect spin transport.
We conclude this paper in Sec.~\ref{sec:conclusion}.

Throughout this paper, we set $\hbar=k_B=1$ and the volumes for both reservoirs are taken to be unity.

\section{\label{sec:formalism}Formalism}
In order to study spin transport of two terminal systems in normal Fermi gases with strong interaction, we begin with the following grand canonical Hamiltonian:
\begin{subequations}\label{eq:Hamiltonian}
\begin{align}
K&=K_R+K_L+H_T,\\
K_j&=\sum_{\p} \sum_{\sigma=\up,\down}\xi_{\p,\sigma,j}c_{\p,\sigma,j}^\+c_{\p,\sigma,j}\nonumber\\
&\quad-U\sum_{\p,\p',\q}c_{\p+\q,\up,j}^\+c_{-\p,\down,j}^\+c_{-\p',\down,j}c_{\p'+\q,\up,j},\\
\label{eq:H_T}
H_T&=t\sum_{\sigma=\up,\down}\sum_{\p,\p'}c_{\p,\sigma,R}^\+c_{\p',\sigma,L}+\mathrm{H.c.},
\end{align}
\end{subequations}
where left and right reservoirs are referred to as $j=L,R$, respectively, $c_{\p,\sigma,j}$ is the annihilation operator of a fermionic atom with spin $\sigma=\,\up,\down$ in the reservoir $j$, and $\xi_{\p,\sigma,j}=\p^2/2m-\mu_{\sigma,j}$ is a single-particle energy measured from the chemical potential $\mu_{\sigma,j}$.
The interaction between fermions in the reservoir $j$ is attractive ($U>0$) and related to the $s$-wave scattering length $a$ by $1/U=\sum_{\p}m/\p^2-m/(4\pi a)$.
The term in $H_T$ denotes the tunneling of fermions from one reservoir to the other and $t$ characterizes the strength of the tunneling.

As mentioned above, we focus on the situation where there is a pure spin bias shown in Fig.~\ref{fig:setup}.
The majority and minority components in the left reservoir are $\up$ and $\down$, respectively, and  those in the right reservoirs are the other way around.
It follows that the parameter $h$ controls not only the polarizations in both reservoirs but also a bias in spin transport.
We assume that both reservoir has same temperature $T$, which is above the superfluid transition temperature.
For convenience, we define a new label $\alpha\equiv(\sigma,j)$ and, hereafter, the majority components $(\sigma,j)=(\up,L),(\down,R)$ are referred to as ``$\alpha=+$,'' while the minority ones $(\sigma,j)=(\down,L),(\up,R)$ as ``$\alpha=-$.''
The chemical potentials of the majority ($\alpha=+$) and minority ($\alpha=-$) are $\mu_\pm=\mu\pm h$ as depicted in Fig.~\ref{fig:setup}.

\begin{figure}[t]
\centering
\includegraphics[width=1.0\columnwidth,clip]{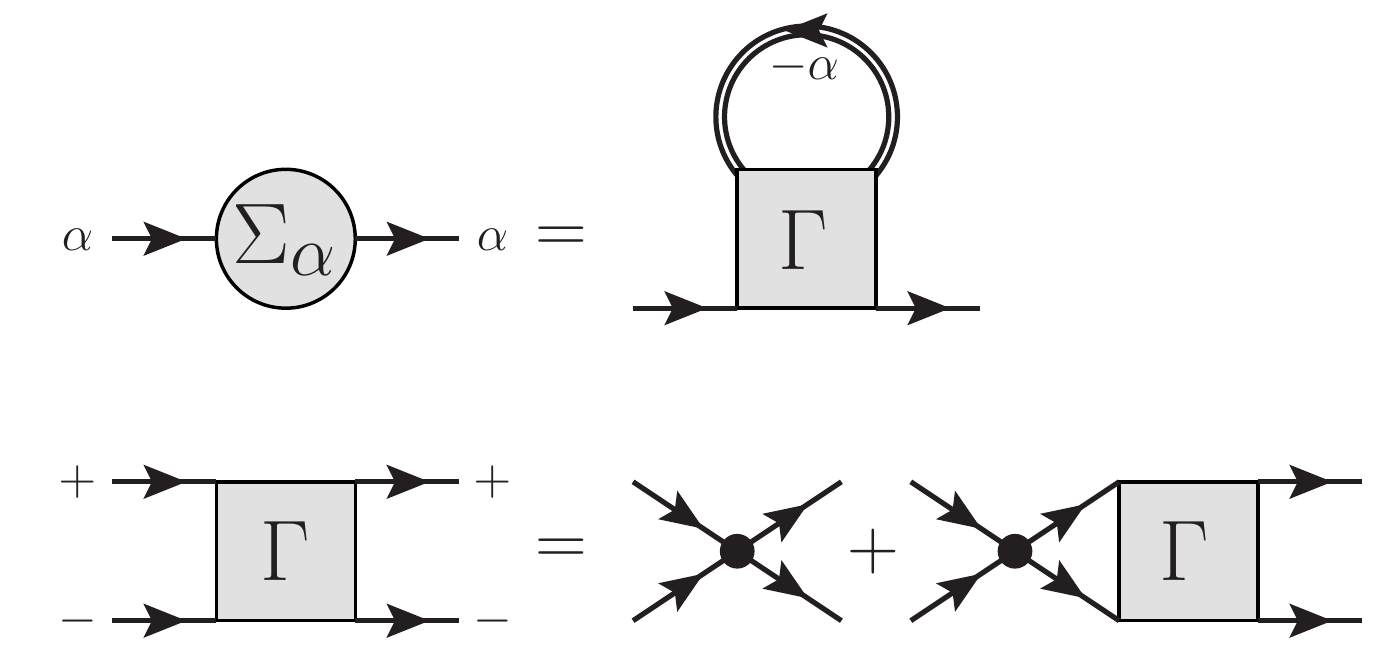}
\caption{\label{fig:feynman}
Feynman diagrams for $\Sigma_\alpha(\p,i\omega_n)$ and $\Gamma(\q,i\nu_\ell)$.
The double (single) line represents the dressed propagator $\mathcal{G}_\alpha(\p,i\omega_n)$ [the bare propagator $1/(i\omega_n-\xi_{\p,\alpha})$] while the dot denotes the interparticle attraction $-U$.
The index $-\alpha$ denotes the component opposite to $\alpha$.}
\end{figure}
The spin current operator in the Heisenberg representation is given by $\hat{I}_\mathrm{spin}(t')=-\dot{N}(t')_{\up,L}+\dot{N}(t')_{\down,L}$, where $N_{\sigma,j}=\sum_{\p}c_{\p,\sigma,j}^\+c_{\p,\sigma,j}$ is the particle number operator.
By using the linear response theory, the spin current to leading order in the tunneling amplitude $t$ can be obtained in a similar way as for a mass current~\cite{Mahan:2013}.
For a steady state, the spin current is obtained as (see Appendix~\ref{appendix:noise})
\begin{align}\label{eq:Ispin}
I_\mathrm{spin}&=4\pi t^2\int\! d\omega\,\rho_+(\omega-h)\rho_-(\omega+h)\nonumber\\
&\quad\quad\quad\times\left[f(\omega-h)-f(\omega+h)\right],
\end{align}
where $\rho_\alpha(\omega)$ is the DOS for the majority ($\alpha=+$) [minority ($\alpha=-$)] and $f(\omega)=1/(e^{\omega/T}+1)$ is the Fermi distribution function.
We note that  regardless of  value of $h$, Eq.~(\ref{eq:Ispin}) is correct up to $t^2$.
In order to take contributions of pair correlations to $\rho_\alpha(\omega)$ into account, we employ the extended $T$-matrix approximation (ETMA)~\cite{Kashimura:2012,Tajima:2018}.
The density of states is related to an analytically-continued  Matsubara Green's function $\mathcal{G}_\alpha(\p,i\omega_n)$ with $\omega_n=(2n+1)\pi\,T$ ($n\in \mathbb{Z}$) as follows:
\begin{align}\label{eq:DOS}
\rho_\alpha(\omega)=-\frac1\pi \sum_{\p}\mathrm{Im}[\mathcal{G}_\alpha(\p,i\omega_n\to\omega+i\delta)],
\end{align}
where
\begin{align}
\mathcal{G}_\alpha(\p,i\omega_n)=\frac{1}{i\omega_n-\xi_{\p,\alpha}-\Sigma_\alpha(\p,i\omega_n)},
\end{align}
$\xi_{\p,\alpha}=\p^2/2m-\mu_\alpha$, and $\delta$ is an infinitesimal positive constant.
Within the ETMA, the self energy $\Sigma_\alpha(\p,i\omega_n)$ and the many-body $T$-matrix $\Gamma(\q,i\nu_\ell)$ with $\nu_\ell=2\ell\pi\, T$ ($\ell\in\mathbb{Z}$) are given by Feynman diagrams in Fig.~\ref{fig:feynman}, leading to
\begin{align}\label{eq:Sigma}
\Sigma_\pm(\p,i\omega_n)&=T\sum_{\q,i\nu_\ell}\Gamma(\q,i\nu_\ell)\mathcal{G}_\mp(\q-\p,i\nu_\ell-i\omega_n),\\
\frac{1}{\Gamma(\q,i\nu_\ell)}&=-\frac1U-\sum_{\p}\frac{1-f(\xi_{\p+\q/2,+})-f(\xi_{-\p+\q/2,-})}{i\nu_\ell-\xi_{\p+\q/2,+}-\xi_{-\p+\q/2,-}}.
\end{align}
We note that the self-consistent programme above can reduce unphysical results.
For example, the ordinary $T$-matrix approximation is known to suffer from a negative spin susceptibility in the strong-coupling regime.
However, ETMA spin susceptibility takes a positive value in the whole crossover regime~\cite{Kashimura:2012}.

\begin{figure}[t]
\centering
\includegraphics[width=0.70\columnwidth,clip]{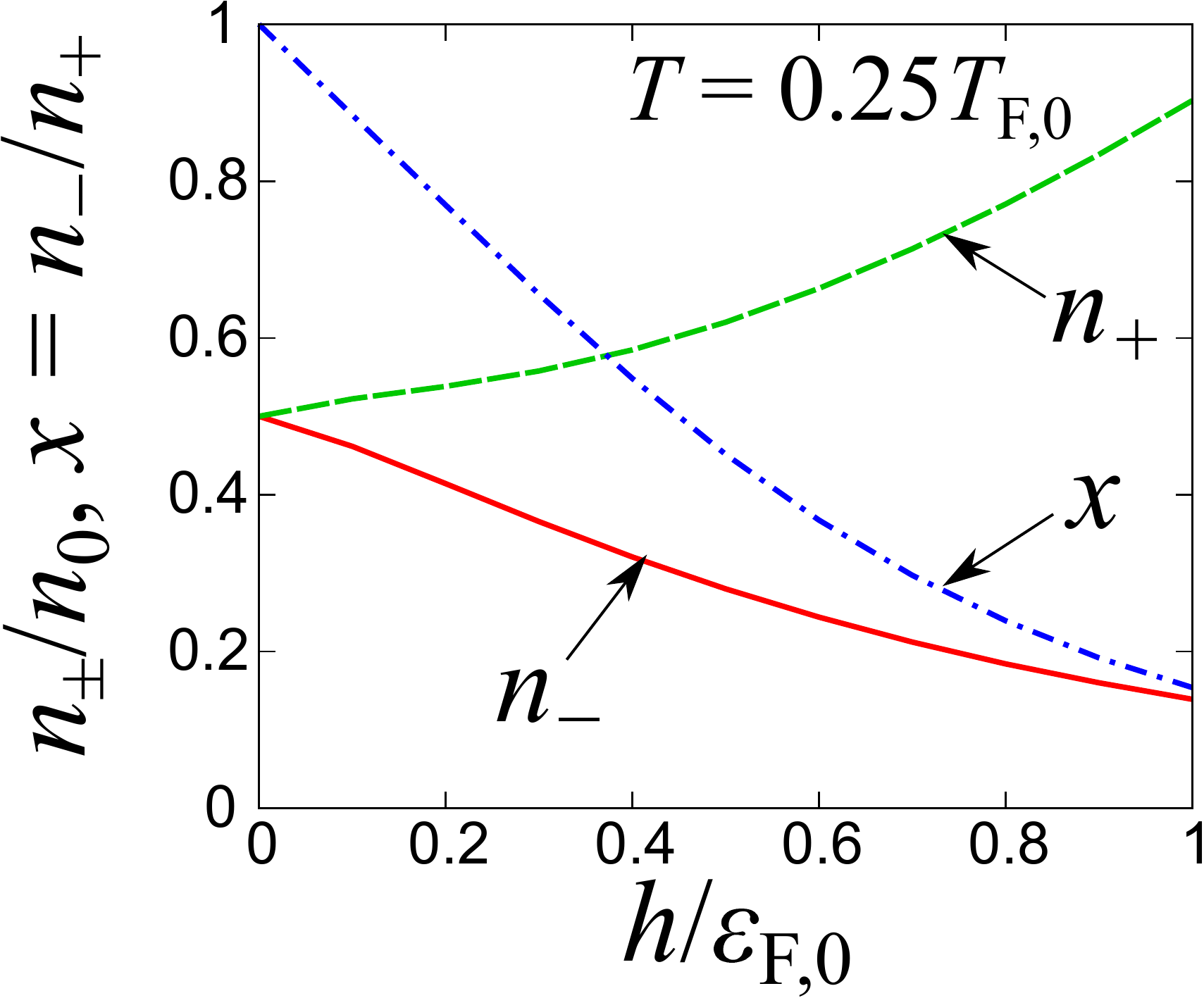}
\caption{\label{fig:density}
Number densities defined by $n_\pm=T\sum_{\p,i\omega_n}\mathcal{G}_\pm(\p,i\omega_n)$ and the ratio $x=n_-/n_+$ at $(k_{\mathrm{F},0} a)^{-1}=0$.
Here, $\varepsilon_{\mathrm{F},0}=T_{\mathrm{F},0}$ is the Fermi energy in the unpolarized case ($h=0$).}
\end{figure}
We now discuss the choice of parameters $(a,T,\mu,h)$ in this work.
In an unpolarized case ($h=0$), we fix $(a,T,\mu)$ as follows.
For a given density $n_0$ of the total particle number in each reservoir, the corresponding Fermi momentum and temperature are provided by $k_{\mathrm{F},0}=(3\pi^2n_0)^{1/3}$ and $T_{\mathrm{F},0}=k_{\mathrm{F},0}^2/(2m)$, respectively, and two dimensionless parameters $(k_{\mathrm{F},0} a)^{-1}$ and $T/T_{\mathrm{F},0}$ are fixed.
Then, $(k_{\mathrm{F},0} a)^{-1}\to-\infty$ corresponds to the weak interaction limit and $(k_{\mathrm{F},0} a)^{-1}\to+\infty$ to the strong interaction limit
in the fermion language.
The spin-averaged chemical potential $\mu$ is determined so that the particle number equation in the absence of $h$,
\begin{align}\label{eq:number}
n_0&=T\sum_{\alpha=\pm}\sum_{\p,i\omega_n}\mathcal{G}_\alpha(\p,i\omega_n)|_{h=0},
\end{align}
is satisfied. 
Then, the fictitious Zeeman field $h$ is changed with $(a,T,\mu)$ fixed.
Since we are interested in how strong correlations affect spin transport, we consider the regime $|k_{\mathrm{F},0}a|^{-1}\lesssim1$ near unitarity.
Figure~\ref{fig:phase} shows the phase diagram of the Fermi gases at unitarity ($(k_{\mathrm{F},0}a)^{-1}=0$) in the $(h,T)$-plane.
The transition temperature is determined as the temperature satisfying the Thouless criterion~\cite{Thouless:1960} given by $[\Gamma(\mathbf{0},0)]^{-1}=0$.
We note that at low temperature there is the first-order phase transition between normal and superfluid phases, including the so-called Fulde-Ferrel-Larkin-Ovchinnikov state~\cite{Sheehy}.
To address these transitions, one has to calculate the thermodynamic potential in each phase, which is out of scope in this paper.
Figure~\ref{fig:density} shows the number densities $n_\pm$ of majority and minority atoms at unitarity as functions of $h$.
(Note that the total number density $n_++n_-$ changes as $h$ increases.)
The monotonic behavior of $x=n_-/n_+$ in Fig.~\ref{fig:density} means that, as $h$ becomes larger, the gases in both reservoirs become more highly polarized.
As shown in Figs.~\ref{fig:spectrum} and \ref{fig:polaronDOS} in the next section, the polaron picture becomes valid for the large $h$ regime.

\section{\label{sec:result}Result}
\begin{figure}[t]
\centering
\includegraphics[width=6cm]{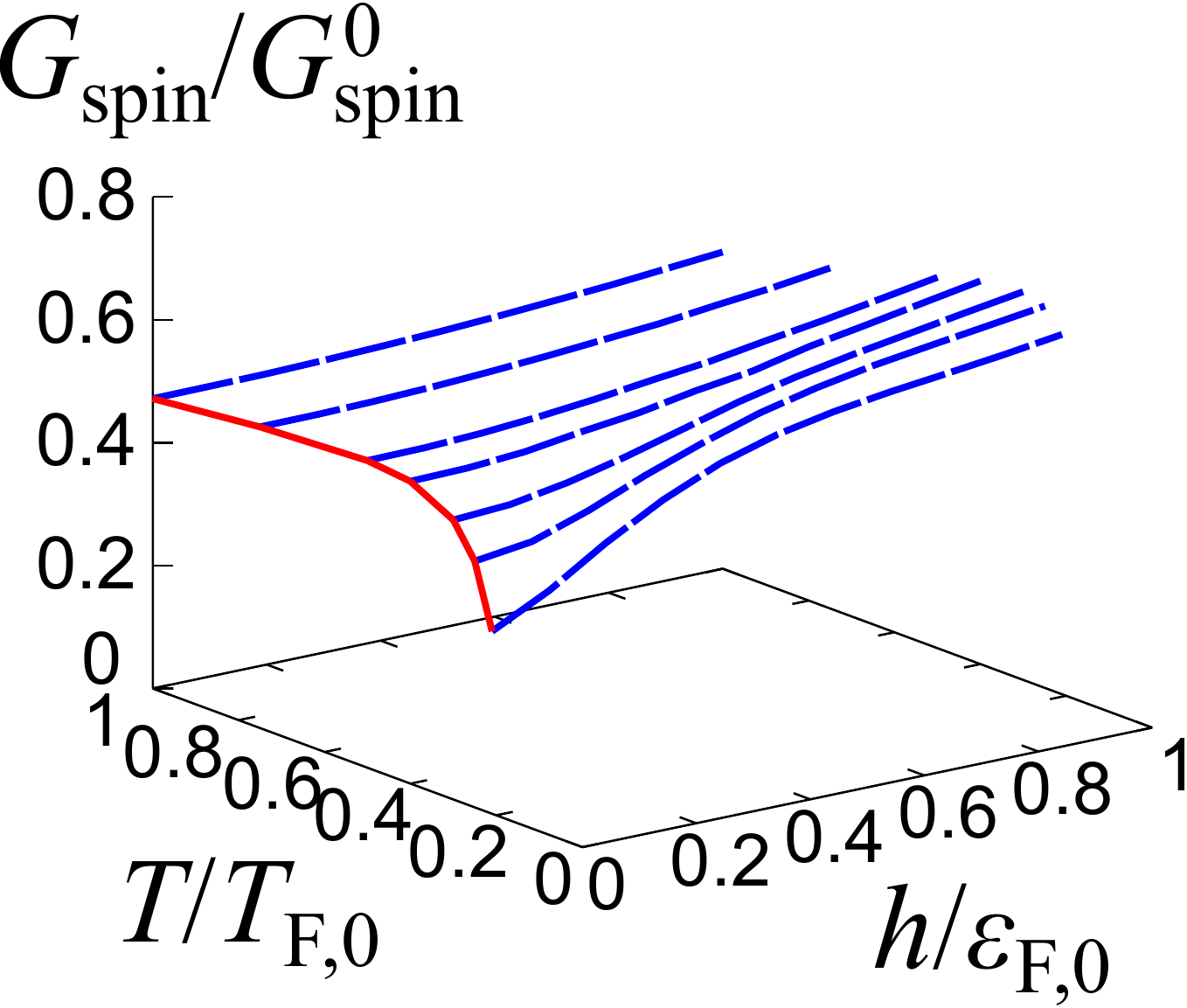}
\caption{\label{fig:Gspin@unitarity}
Spin conductance $G_\mathrm{spin}$ at unitarity.
Here, $G_\mathrm{spin}^{0}=2m^3t^2\varepsilon_{\mathrm{F},0}/\pi^3$ is the zero-bias conductance at $T=0$ in the absence of interactions (see Appendix~\ref{appendix:free}).
The red solid line denotes $G_\mathrm{spin}$ in the zero-bias limit.}
\end{figure}
We now discuss spin transport properties for normal Fermi gases with strong interparticle interactions under the configuration shown in Fig.~\ref{fig:setup}.
Figure~\ref{fig:Gspin@unitarity} shows the spin conductance $G_\mathrm{spin}\equiv I_\mathrm{spin}/(2h)$ at unitarity.
We can see that $G_\mathrm{spin}$ grows with increasing $h$ with a fixed $T$.
Furthermore, it is remarkable that $G_\mathrm{spin}$ is largely suppressed in the low-$T$ and low-$h$ regime, where the pseudogap emerges.
\begin{figure}[t]
\centering
\includegraphics[width=6cm]{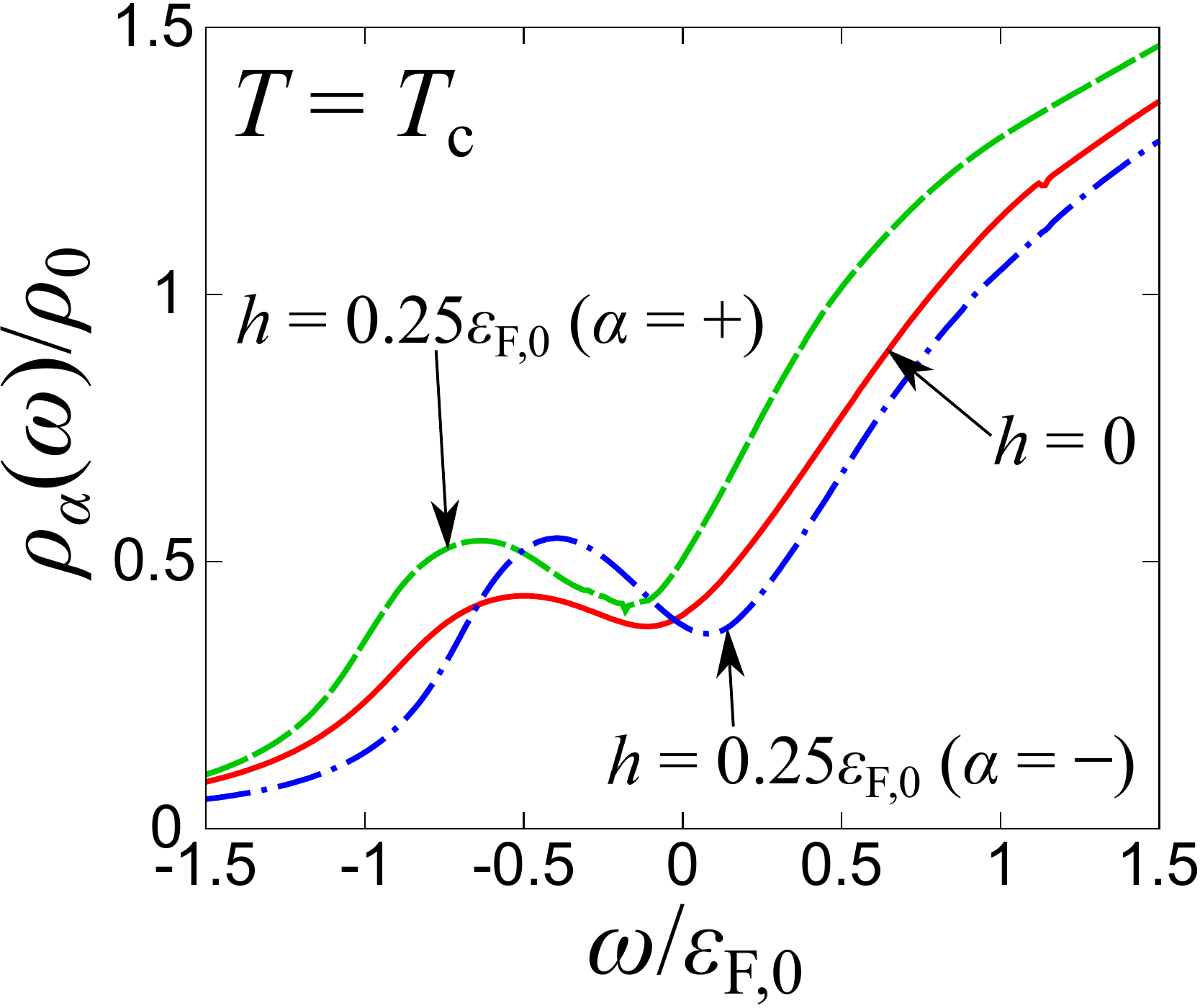}
\caption{\label{fig:pseudogap}Densities of states at critical temperature $T_{\rm c}$ in the unitary limit $(k_{\mathrm{F},0}a)^{-1}=0$.
Here, $\rho_0=m\sqrt{2m\varepsilon_{\mathrm{F},0}}/(2\pi^2)$ is the density of states for an unpolarized free Fermi gas at the Fermi energy $\varepsilon_{\mathrm{F},0}$.
}
\end{figure}
As shown in Fig.~\ref{fig:pseudogap},
our ETMA calculation  for unpolarized ($h=0$) and polarized ($h=0.25\varepsilon_{\mathrm{F},0}$) gases at $T=T_\mathrm{c}$
can  confirm the pseudogap structures of DOSs, being  the signature of the preformed Cooper pairs in the BCS-BEC crossover regime of an ultracold Fermi gas.

\begin{figure}[t]
\centering
\includegraphics[width=6cm]{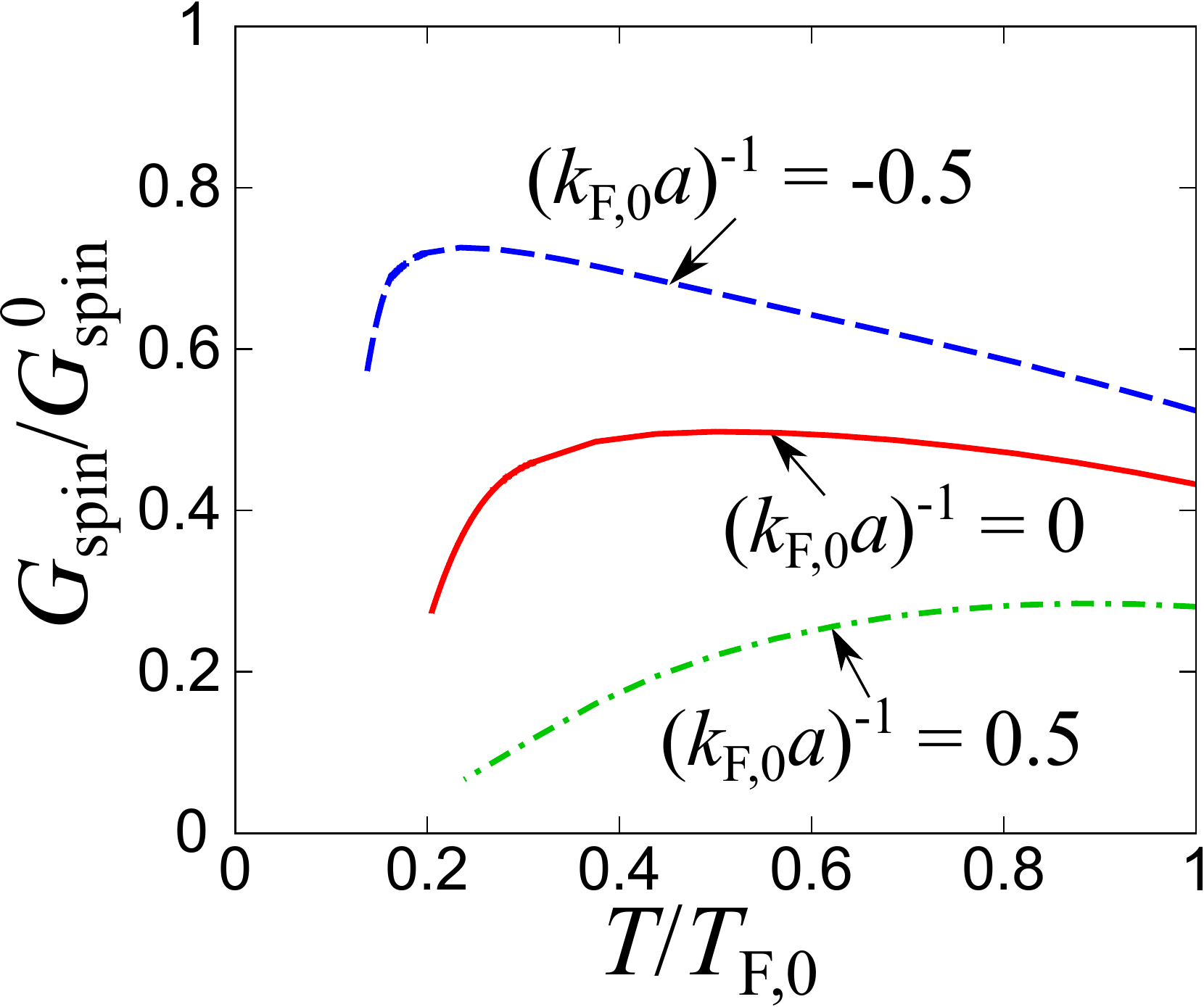}
\caption{\label{fig:Gspin}
Spin conductance $G_\mathrm{spin}$ in the zero-bias limit.
The endpoint of each line corresponds to $G_\mathrm{spin}$ at $T=T_\mathrm{c}$.}
\end{figure}

To make the effect of the pseudogap on spin transport clearer, let us focus on the zero-bias limit.
In this limit, majority and minority DOSs become identical, $\bar{\rho}(\omega)=\rho_\pm(\omega)|_{h=0}$, and the spin conductance reduces to the following form:
\begin{align}\label{eq:G_spin}
G_\mathrm{spin}&=4\pi t^2\!\!\int\!\!d\omega\,[\bar{\rho}(\omega)]^2\!\left(-\frac{\partial f(\omega)}{\partial \omega}\right)\ \ \text{for } h\to0.
\end{align}
Figure~\ref{fig:Gspin} provides the obtained temperature dependence of $G_\mathrm{spin}$ at $(k_{\mathrm{F},0}a)^{-1}=-0.5,\, 0,\, 0.5$, in the zero-bias limit.
We note that the calculation of $G_\mathrm{spin}$ is stopped at the transition temperature $T_\mathrm{c}$.
Away from $T_\mathrm{c}$, $G_\mathrm{spin}$ increases with decreasing $T$ because of quantum degeneracy of fermions.
On the other hand, as shown in Fig.~\ref{fig:pseudogap}, the DOS has a dip structure around $\omega=0$ near the superfluid transition.
Since $-\partial f(\omega)/(\partial \omega)\propto\cosh^{-2}(\omega/2T)$, the spin conductance is sensitive to $\bar{\rho}(\omega)$ around $\omega=0$.
Therefore, the appearance of this pseudogap leads to a large suppression of $G_\mathrm{spin}$.
 
 The single-particle excitation is strongly suppressed due to the formation of spin-singlet pairs in the pseudogap regime. This suppression leads to the so-called spin gap in the temperature dependence of the spin susceptibility~\cite{Palestini:2012,Mink:2012,Enss:2012,Wlazlowski:2013,Tajima:2014,Jensen:2019}.
We note that the spin-gap temperature, where the spin susceptibility starts to be suppressed due to strong pairing fluctuations, is $T_{\rm SG}=0.37T_{\mathrm{F},0}$ at unitarity~\cite{Tajima:2014}.
Although it is the crossover temperature and there are ambiguities for the definition to characterize these phenomena, the maximum temperature of $G_{\mathrm{spin}}$ at unitarity is also close to $T_{\rm SG}$.
This result indicates that $G_{\mathrm{spin}}$ is also useful to study pseudogap physics.
From Fig.~\ref{fig:Gspin}, we can also see the interaction dependence of $G_\mathrm{spin}$.
In the weak-coupling side ($(k_{\mathrm{F},0}a)^{-1}=-0.5$ in Fig.~\ref{fig:Gspin}), $G_{\mathrm{spin}}$ becomes larger compared to that at unitarity.
However, even at this coupling, $G_{\mathrm{spin}}$ near $T_{\rm c}$ decreases due to the pairing fluctuation effects as decreasing temperature.
As in the case of the spin susceptibility~\cite{Tajima:2014}, the spin conductance is strongly suppressed as increasing strength of the pairing interaction.
At stronger couplings, the gases in both reservoirs are dominated by tightly bound molecules and the spin degree of freedom tends to be frozen.
In this case, only the thermally dissociated atoms contribute to the spin susceptibility as well as spin transport.

\begin{figure}[t]
\centering
\includegraphics[width=6cm]{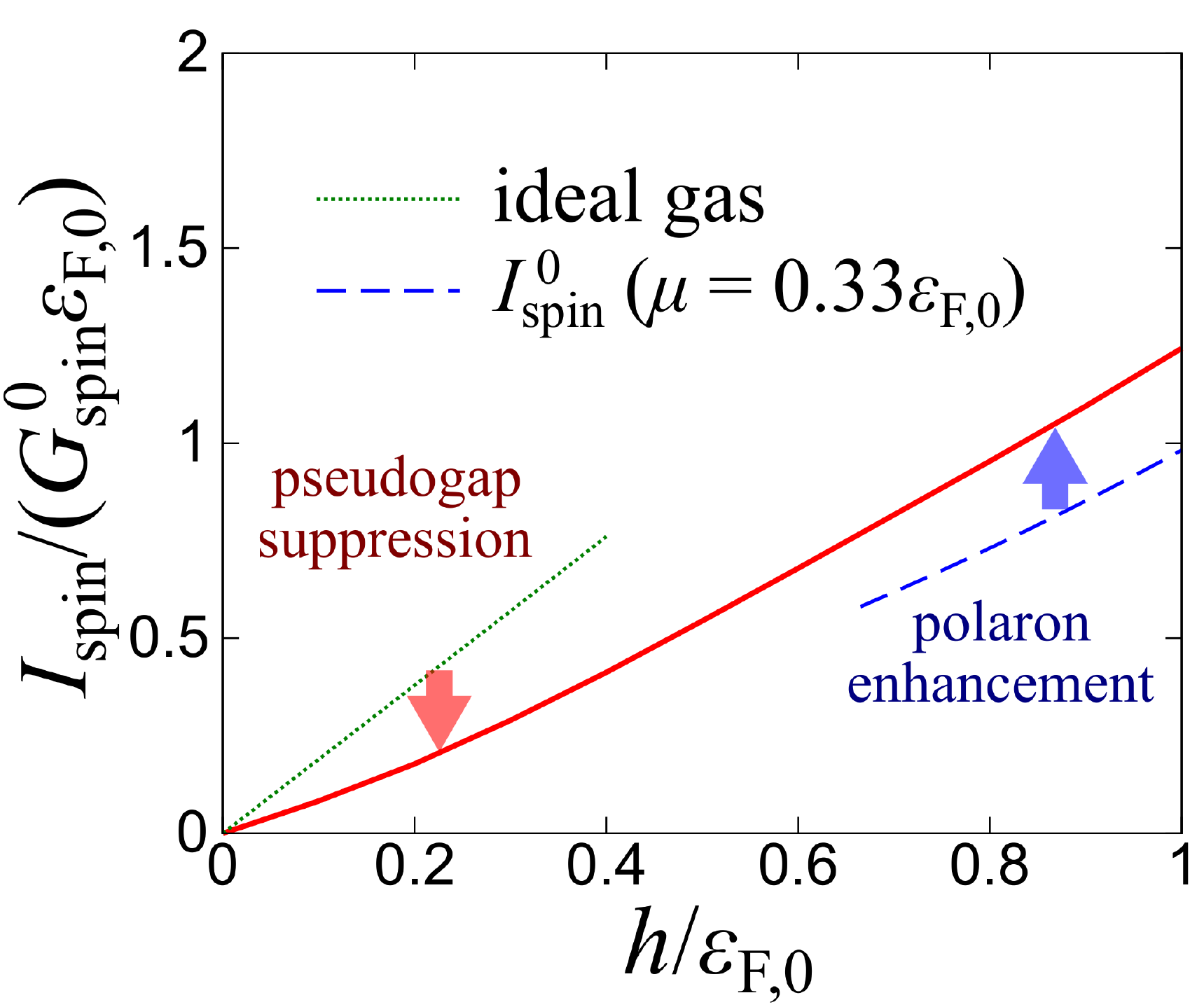}
\caption{\label{fig:Ispin@unitarity}
Bias dependence of $I_\mathrm{spin}$ (solid line) at $T=0.25T_{\mathrm{F},0}$ in the unitary limit.
The green dotted line denotes the current in the non-interacting case, 
while the blue dashed line denotes the current without self-energy corrections to $\rho_\pm(\omega)$~\cite{note}.}
\end{figure}
\begin{figure}[t]
\centering
\includegraphics[width=6cm]{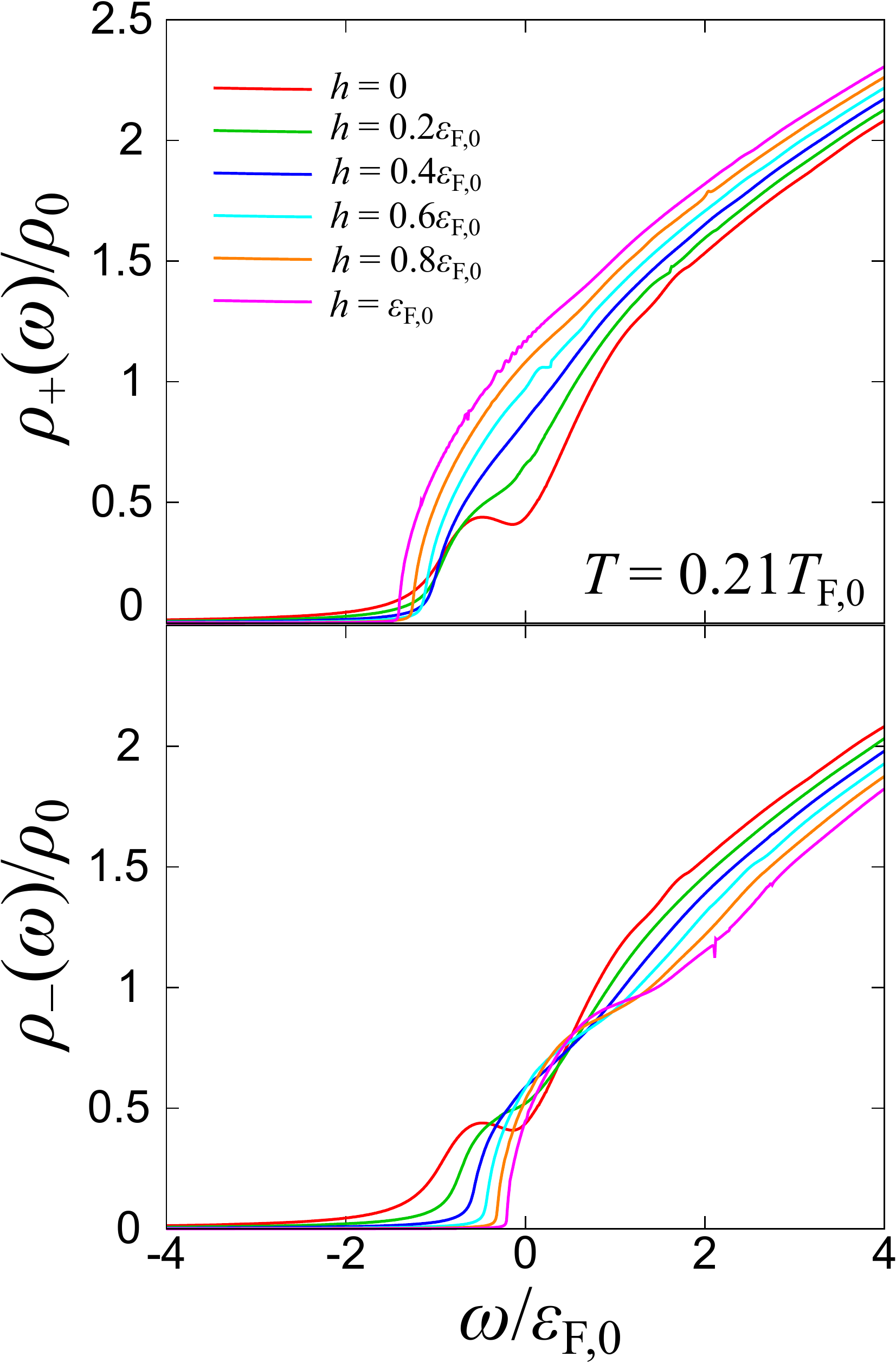}
\caption{\label{fig:DOSs}
Majority and minority DOSs at $T=0.21T_{\mathrm{F},0}$ in the unitary limit for various $h$.}
\end{figure}

Figure~\ref{fig:Ispin@unitarity} represents a crossover of the spin current at unitarity from the pseudogap regime to the polaronic regime by shifting $h$. 
We can see that $I_\mathrm{spin}$ for small $h$ is smaller than in the non-interacting counterpart, where the current is analytically given by Eq.~(\ref{eq:I_spin^(0)}) in Appendix~\ref{appendix:free}.
As explained in the discussion of $G_\mathrm{spin}$, this suppression is caused by the pseudogap in the region where $h$ is small.
Figure~\ref{fig:DOSs} shows the calculated DOSs for both majority and minority components at $T=0.21T_{\mathrm{F},0}$ for various $h$.
When $h$ becomes larger, polarizations of the gases in both reservoirs grow and the pseudogap structures of $\rho_\pm(\omega)$ vanish since the gases go away from $T_\mathrm{c}$ at a fixed temperature.
The majority DOS is enhanced in the whole energy region with increasing $h$ due to the increase of $n_+$.
In the large-$h$ region, $\rho_+(\omega)$ coincides with a DOS in an ideal Fermi gas given by Eq.~(\ref{freeDOS}), since under a large population imbalance, minority atoms cause a negligible effect to a large number of majority atoms.
On the other hand, the minority DOS shows a more complex modification than $\rho_+(\omega)$ with increasing $h$.
In particular, in the large-$h$ regime, minority atoms can be regarded as the so-called Fermi polarons.
In our configuration of chemical potentials with fixed $\mu=(\mu_++\mu_-)/2$, $I_\mathrm{spin}$ is enhanced compared with that without self-energy corrections~\cite{note}.
This implies that the polaronic quasi-particle excitations encoded in the self-energy corrections play an important role in spin transport for a large spin bias.

\begin{figure}[t]
\centering
\includegraphics[width=7cm]{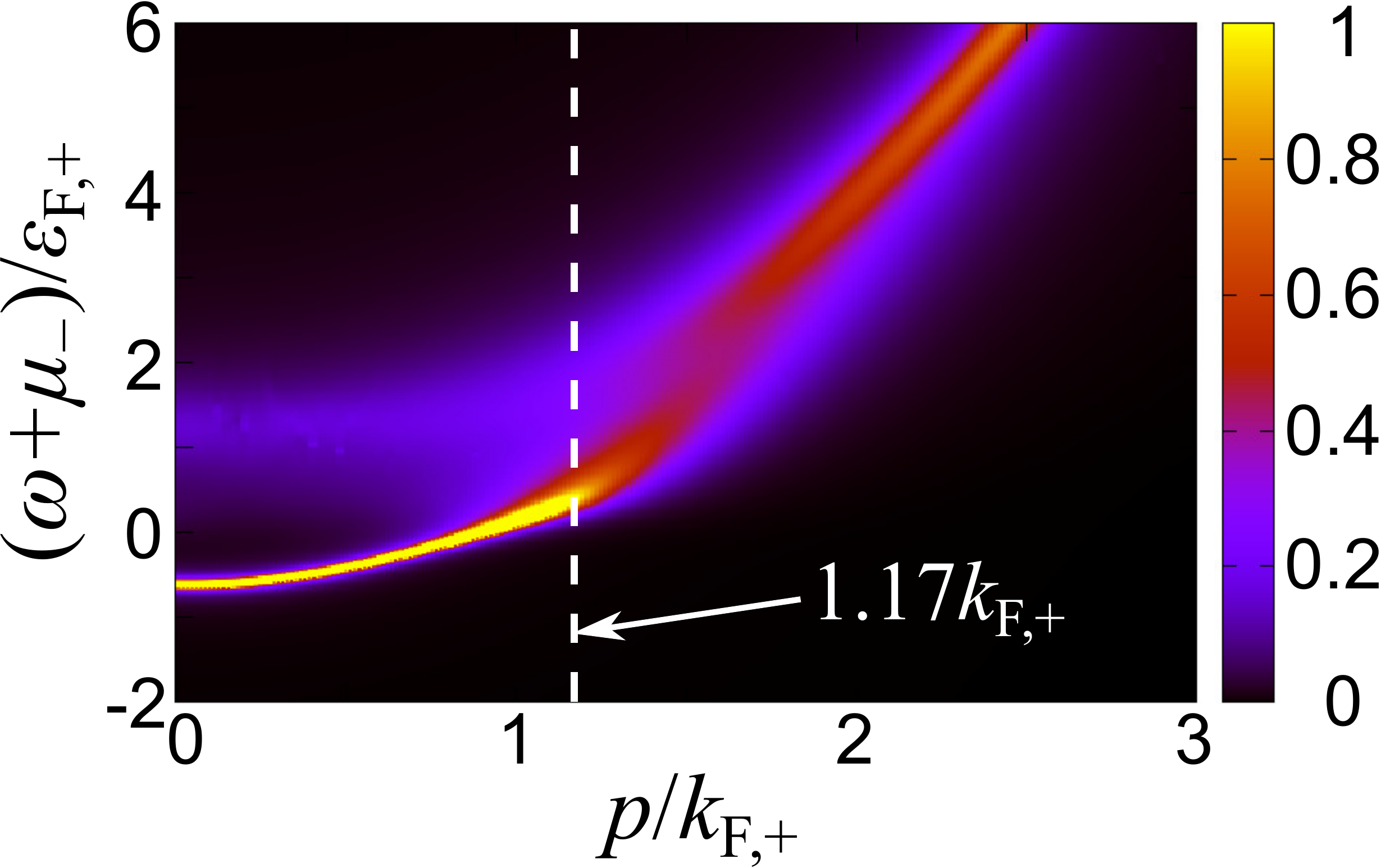}
\caption{\label{fig:spectrum}
Minority spectral function $A_-(\p,\omega)$ (arbitrary unit) at $T=0.1T_{\mathrm{F},+}$ for $n_-/n_+=0.1$ in the unitary limit.
The vertical dashed line shows $p=1.17k_{\mathrm{F},+}$ where the velocity of an attractive polaron reaches the Fermi velocity of medium atoms.
Note that $k_{\mathrm{F},+}$ and $\varepsilon_{\mathrm{F},+}=T_{\mathrm{F},+}$ are the Fermi momentum and the Fermi energy for the majority component, respectively.}
\end{figure}
In order to discuss the contributions of the polaronic transport to $I_\mathrm{spin}$ for large $h$, we start with the investigation of the single-particle spectral functions defined by 
\begin{align}
A_\alpha(\p,\omega)&=-\frac{1}{\pi}\mathrm{Im}[\mathcal{G}_\alpha(\p,\omega+i\delta)].
\end{align}
In the literature on the Fermi polarons, the Fermi momentum $k_{\mathrm{F},+}=(6\pi^2n_+)^{1/3}$ and the Fermi energy $\varepsilon_{\mathrm{F},+}=k_{\mathrm{F},+}^2/(2m)$ for majority atoms are conventionally taken as units of energy and momentum.
Thus we use these units to discuss $A_\alpha(\p,\omega)$ in the large spin-bias regime.
For large $h$, the majority spectral function can be replaced by that for free fermions $A_+(\p,\omega)=\delta(\omega+\mu_+-\frac{p^2}{2m})$.
On the other hand, Fig.~\ref{fig:spectrum} shows that the minority spectral functions near unitarity are distinct from its non-interacting counterpart.
In particular, there are two types of characteristic excitations: a sharp peak at low energy corresponding to an attractive polaron and a broad peak around $\omega+\mu_-\approx\varepsilon_{\mathrm{F},+}$ associated with a repulsive branch or repulsive polarons.
Since $A_-(\p,\omega)$ at low energy is dominated by the attractive polaron, it is well approximated by 
\begin{align}
A_-(\p,\omega)&\approx Z_a\delta\left(\omega+\mu_--\frac{p^2}{2m_a^*}-E_a\right),
\end{align}
where $Z_a$, $m_a^*$, and $E_a$ are the renormalization factor, the effective mass, and the energy of the attractive polaron, respectively.
By integrating this approximated form over $\p$, the contribution from the polaron to $\rho_-(\omega)$ is found to be
\begin{align}\label{eq:polaronDOS}
\rho_-^{\mathrm{(att)}}(\omega)&=\frac{m_a^*Z_a}{2\pi^2}\sqrt{2m_a^*(\omega+\mu_--E_a)}
\end{align}
for $\omega>-\mu_-+E_a$.

\begin{figure}[t]
\centering
\includegraphics[width=6cm]{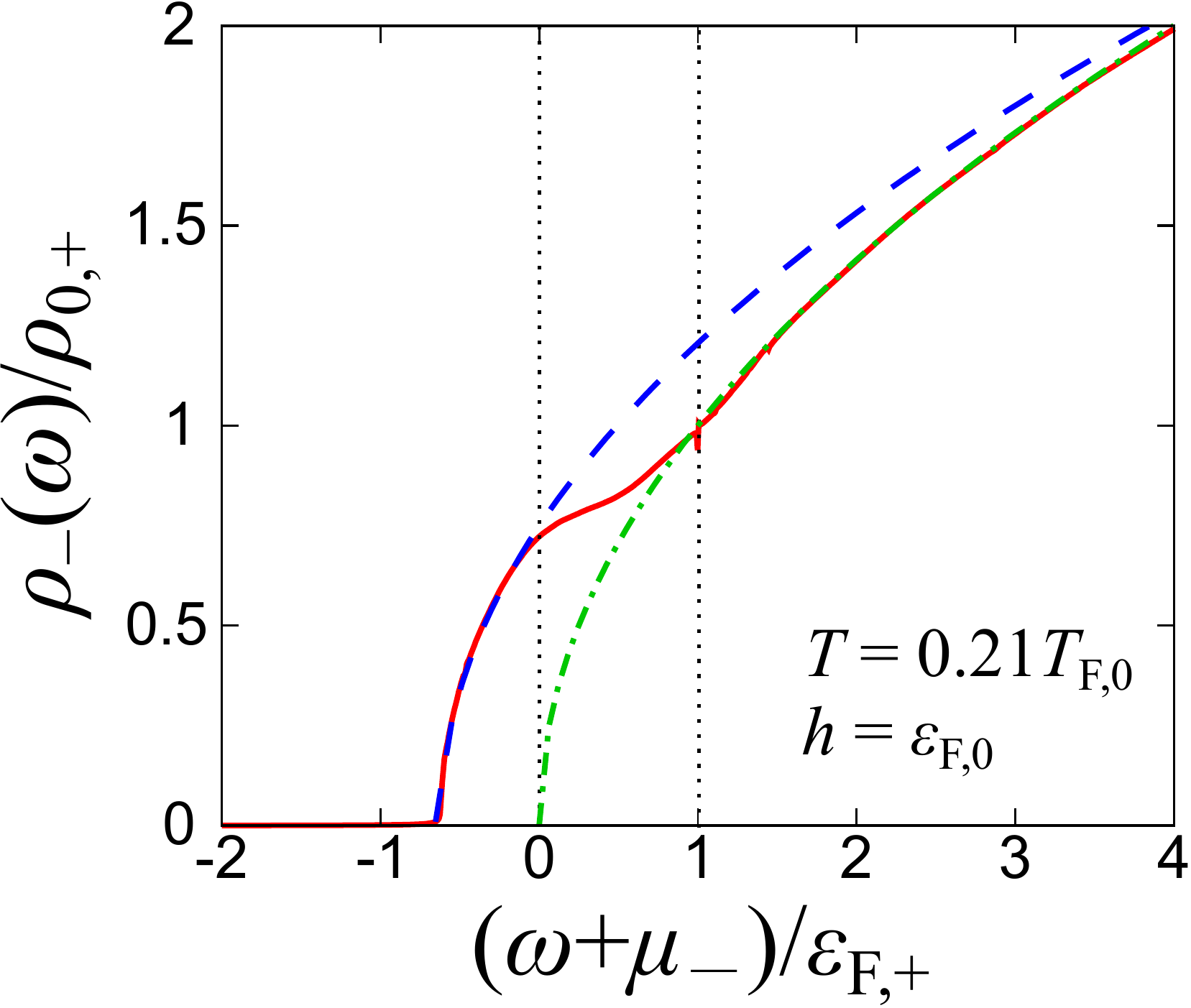}
\caption{\label{fig:polaronDOS}
Minority DOS for a highly polarized gas at unitarity.
The blue dashed line denotes the fitting curve by using Eq. (\ref{eq:polaronDOS}), while the green chain line denotes a non-interacting DOS in Eq.~(\ref{freeDOS}).
For usual convention in polaron studies, we use the majority Fermi energy $\varepsilon_{\mathrm{F},+}=(6\pi^2 n_+)^{2/3}/(2m)$ and a non-interacting majority DOS $\rho_{0,+}=m\sqrt{2m\varepsilon_{\mathrm{F},+}}/(2\pi^2)$ at the Fermi energy $\varepsilon_{\mathrm{F},+}$.}
\end{figure}
Figure~\ref{fig:polaronDOS} shows the minority DOS at unitarity numerically calculated in the ETMA.
The obtained $\rho_-(\omega)$ for $\omega+\mu_-\lesssim\varepsilon_{\mathrm{F},+}$ is found to be enhanced compared to a DOS without self-energy corrections.
Fitting the calculated DOS at low energy by using Eq.~(\ref{eq:polaronDOS}), we obtain $Z_a(m_a^*/m)^{3/2}=0.945$ and $E_a=-0.627\varepsilon_{\mathrm{F},+}$.
These results are in good agreement with the zero-temperature results in the single-polaron limit, where $Z_a=0.78$, $m_a^*/m=1.17$, and $E_a=-0.606\varepsilon_{\mathrm{F},+}$, leading to $Z_a(m_a^*/m)^{3/2}=0.987$~\cite{Combescot:2007,Punk:2009}.
This means that the effects of finite $T$ and $x$ on $\rho_-(\omega)$ are not so important in this temperature and bias range~\cite{Tajima:2018}.
For $0\lesssim\omega+\mu_-\lesssim\varepsilon_{\mathrm{F},+}$, $\rho_-(\omega)$ deviates from Eq.~(\ref{eq:polaronDOS}).
In this energy range, the enhancement of $\rho_-(\omega)$ comes from not only the attractive polaron at high momentum, whose lifetime is finite, but also the repulsive branch.
In particular, the broadening of the peak associated with the attractive polaron in Fig.~\ref{fig:spectrum} can intuitively be understood as follows.
The polaron is a quasiparticle consisting of a minority atom surrounded by majority atoms.
This quasiparticle picture is valid when the velocity of a dressed minority atom $v_-\approx p/m_a^*$ is smaller than a typical velocity of the majority atoms $v_{\mathrm{F},+}=k_{\mathrm{F},+}/m$.
When $v_-\gesim v_{\mathrm{F},+}$, the majority atoms can no longer follow the fast-moving minority atom and thus the attractive polaron tends to be unstable.
At unitarity, this unstable regime is estimated as $p\gesim(m_a^*/m)k_{\mathrm{F},+}\approx1.17k_{\mathrm{F},+}$ and is consistent with the region where the peak associated with the attractive polaron becomes broad (see Fig.~\ref{fig:spectrum}).
This mechanism of the broadening is analogous to the Cherenkov instability in Bose polarons~\cite{Grusdt:2018,Nielsen:2019}, where a minority atom undergoes the supersonic regime associated with the Bogoliubov phonons. 
Actually, this intermediate energy range of polaron spectra plays a significant role to understand polaronic spin transport as discussed below.
For sufficiently large $\omega$, $\rho_-(\omega)$ becomes close to its non-interacting counterpart, which is consistent with the asymptotic behavior derived by the operator product expansion~\cite{Nishida:2012}.

\begin{figure*}[t]
\centering
\begin{tabular}{c}
\begin{minipage}{0.5\hsize}
\centering
\includegraphics[width=0.8\hsize]{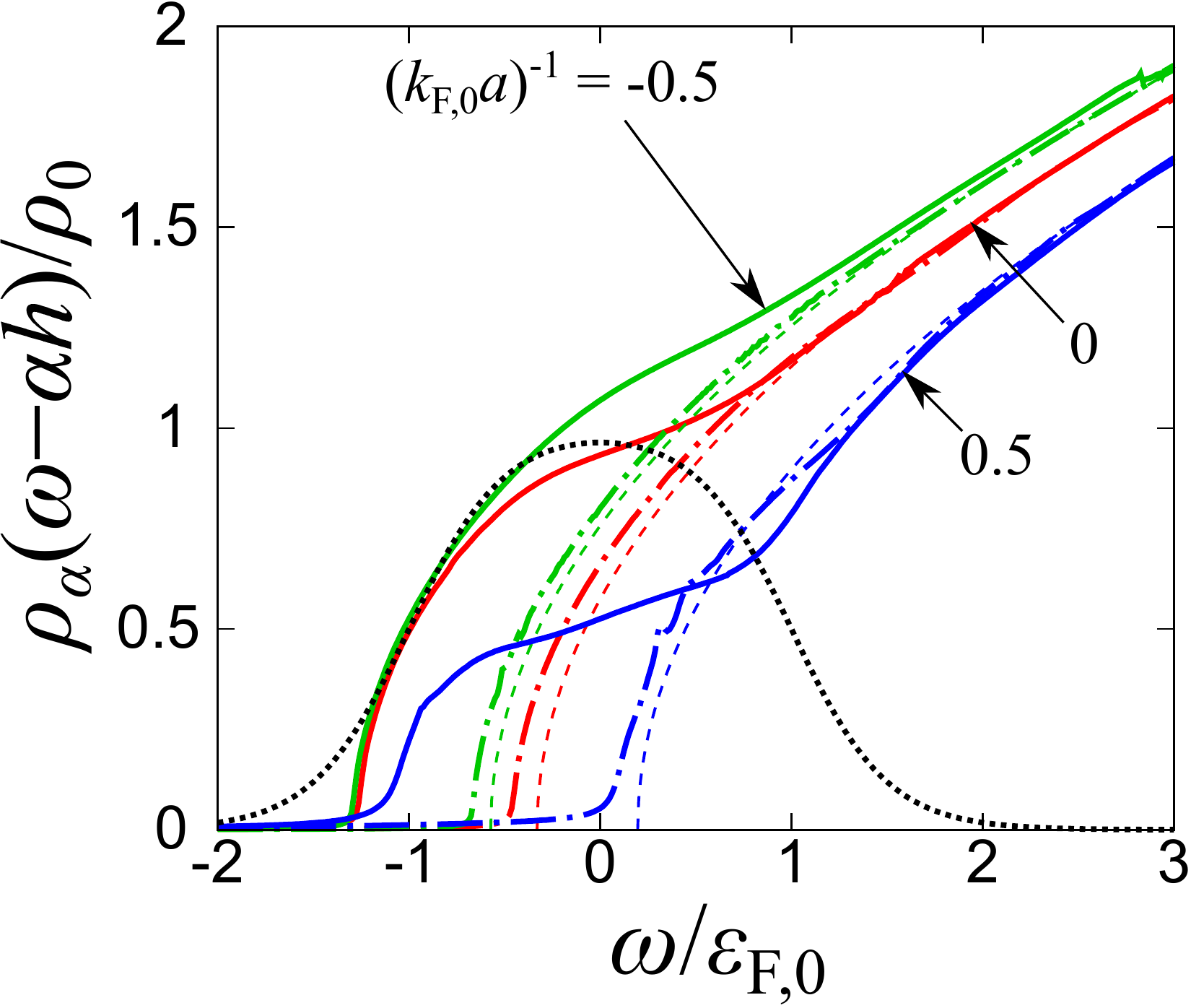}
\hspace{1.6cm}
(a)
\end{minipage}
\begin{minipage}{0.5\hsize}
\centering
\includegraphics[width=0.8\hsize]{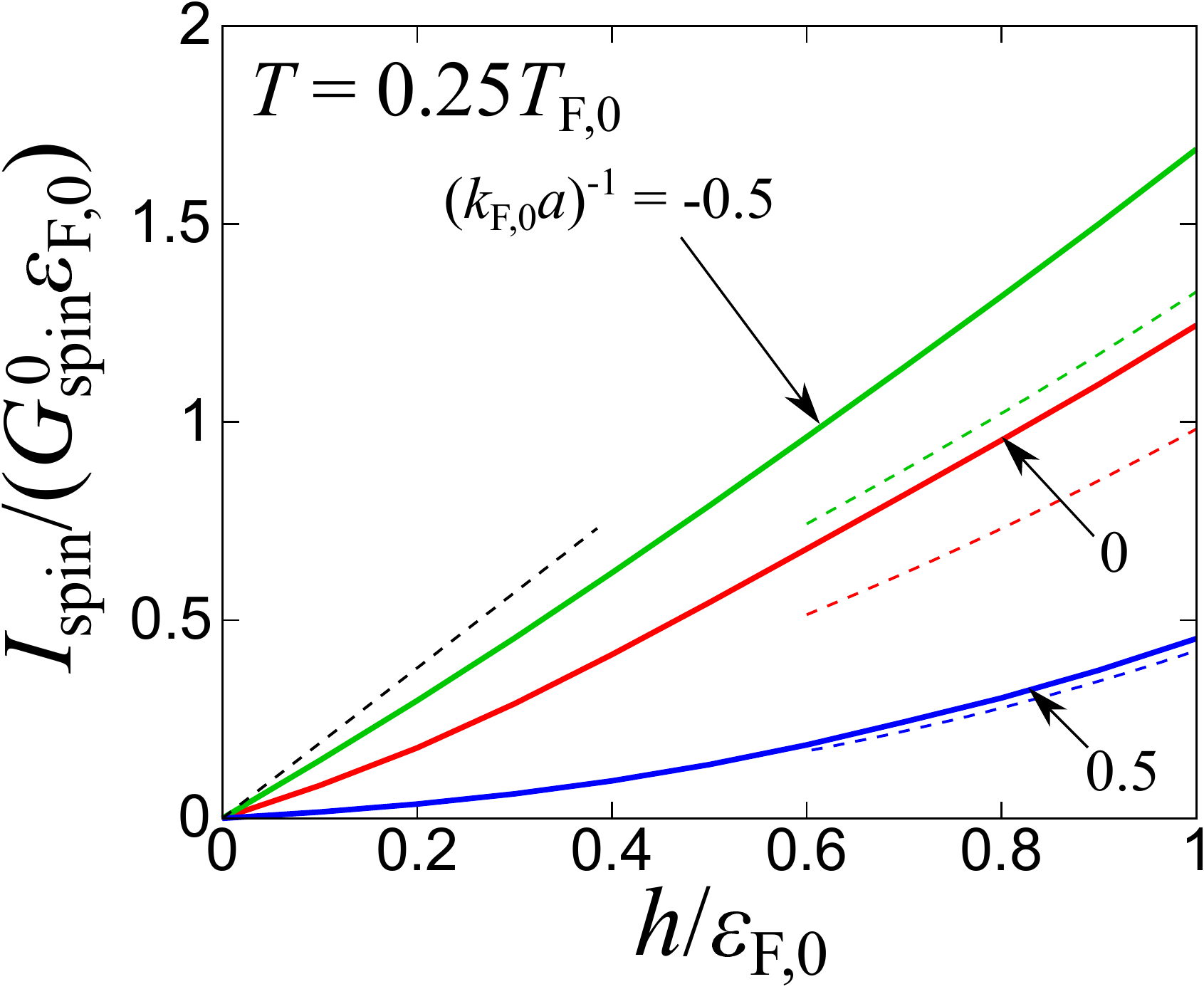}
\hspace{1.6cm}
(b)
\end{minipage}
\end{tabular}
\caption{\label{fig:polaron_transport}
(a) Densities of states in the integrand of Eq.~(\ref{eq:Ispin}) at $T=0.25T_{\mathrm{F},0}$ and $h=\varepsilon_{\mathrm{F},0}$.
The solid lines (chain lines) show the minority (majority) DOSs and the thin dashed lines show the DOSs $\rho_\pm(\omega\mp h)=m\sqrt{2m(\omega+\mu)}/(2\pi^2)$ without interaction corrections.
Note that, for each color, the three lines have the same $\mu$.
The black dotted curve shows $f(\omega-h)-f(\omega+h)$.
(b)
Spin currents at $(k_{\mathrm{F},0}a)^{-1}=-0.5$, $0$, and $0.5$.
The dashed lines show currents without self-energy corrections to the DOSs at each interaction strength.}
\end{figure*}
Let us now come back to $I_\mathrm{spin}$ for a large spin bias.
Figure~\ref{fig:polaron_transport}(a) shows functions in the integrand of Eq.~(\ref{eq:Ispin}) at $T=0.25T_{\mathrm{F},0}$ and $h=\varepsilon_{\mathrm{F},0}$.
We can see that $\rho_+(\omega-h)$ is almost consistent with that without self-energy corrections.
Because of the existences of $f(\omega-h)-f(\omega+h)$ reflecting the Fermi-Dirac statistics and of a threshold energy $\omega=-\mu$ for $\rho_+(\omega-h)$, $\rho_-(\omega+h)$ contributes to $I_\mathrm{spin}$ only over the region where the attractive polaron at finite momentum and the repulsive branch appear.
We note that in the absence of the pairing interaction, both curves coincide with each other.
Thus, the enhancement of the spin current in the highly polarized regime
originates from the polaron excitations in the intermediate-energy range ($0\lesssim \omega+\mu_{-}\lesssim \varepsilon_{\mathrm{F},+}$ in Fig.~\ref{fig:polaronDOS}).

Figure~\ref{fig:polaron_transport}(b) compares $I_\mathrm{spin}$ with and without self-energy corrections at $(k_{\mathrm{F},0}a)^{-1}=-0.5$, $0$, and $0.5$.
While the enhancement of $I_{\mathrm{spin}}$ is  large in the weak-coupling side, it is small in the strong-coupling side.
Since the magnitude of $I_{\mathrm{spin}}$ depends on $\mu$ obtained by solving Eq.~(\ref{eq:number}) at each $(k_{\mathrm{F},0}a)^{-1}$ in our configuration,
$I_{\mathrm{spin}}$ becomes larger in the weak-coupling side where $\mu$ is positively large compared to that in the strong-coupling side.
In this sense, the polaron properties appear as the ratio given by $I_{\mathrm{spin}}$ with and without the self-energy corrections.
This ratio is still larger in the weak-coupling side.
The enhancement of $I_{\mathrm{spin}}$ is associated with  the overlap of $\rho_{+}(\omega+h)$ and $\rho_{-}(\omega-h)$ around $\omega=0$ shown in Fig.~\ref{fig:polaron_transport}(a).
While these two DOSs in the unitary limit as well as the weak-coupling side have a relatively large overlap due to the small polaron energy, such an overlap becomes smaller in the strong-coupling side due to the large polaron energy.
Physically, the large polaron energy indicates the strong binding of Fermi polarons.
Since there is no single-particle state of majority atoms in the energy range corresponding to low-energy attractive polarons, these polaronic states are irrelevant to spin transport.
While the minority DOS at $(k_{\mathrm{F},0}a)^{-1}=0.5$ is enhanced in the low-energy region ($\omega<0$) due to the polaron binding effect,
that around the energy region where majority DOS $\rho_{+}(\omega+h)$ starts to be finite ($\omega\gesim 0.2\varepsilon_{\mathrm{F},+}$) is relatively insensitive to the interaction.
In addition, the strong attraction makes $Z_a$ of the attractive polaron smaller compared with that at unitarity~\cite{Punk:2009}, which is expected to reduce the enhancement of spin transport.
Therefore, one can find that the non-equilibrium spin transport in highly polarized regime is enhanced by the  broadened polaron spectra in the intermediate energy region, and suppressed by the polaron binding effect.

\section{\label{sec:conclusion}Conclusion}
In this paper, we elucidated mesoscopic transport properties of spins for strongly interacting Fermi gases  connected via a quantum point contact.
The tunneling Hamiltonian formalism was used to investigate a steady spin current between two spin-polarized Fermi gases.
By employing the linear response theory combined with the diagrammatic approach, the spin current $I_\mathrm{spin}$ and spin conductance $G_\mathrm{spin}$ in the normal phase were computed for a wide range of parameters.
We found that the emergence of the pseudogap results in a large suppression of spin transport in low-$T$ and low-$h$ regime as shown in Figs.~\ref{fig:Gspin@unitarity}--\ref{fig:Ispin@unitarity}.
On the other hand,the gases become highly polarized for a large spin bias $h$.
In this case, both the attractive polarons at finite momenta and the repulsive branch play significant roles and they lead to the enhancement of $I_\mathrm{spin}$ compared with the current without self-energy corrections.

As mentioned in previous studies on the spin susceptibility for a strongly interacting Fermi gas~\cite{Kashimura:2012,Palestini:2012,Mink:2012,Enss:2012,Wlazlowski:2013,Tajima:2014,Tajima:2016,Tajima:2017,Jensen:2018}, spin properties are sensitive to the formation of a pseudogap.
Mesoscopic spin transport with a small spin bias studied here provides a new probe to experimentally examine the pseudogap phenomenon.
We also clarified that the spin current for a large spin bias is affected by excitations including both the attractive polarons at finite momenta and the repulsive branch.
At the same time, a set of chemical potentials $\mu_{\sigma,j}$ considered in this article (see Fig.~\ref{fig:setup}) has limitation to examine the properties of polarons in the whole energy range.
Such polaronic properties are considered to be accessible in a two terminal system under another choice of $\mu_{\sigma,j}$ with the use of the spin filter, which has been recently realized in an ultracold atom experiment~\cite{Lebrat:2019}.
By assuming $\mu_{\down,L}=\mu_{\down,R}\gg\mu_{\up,L}=\mu_{\up,R}+V_\up$ and filtering out the majority ($\sigma=\down$) component, the fully polarized current for a small bias $V_\up$ encodes information of attractive polarons at low energy.

Our method can be generalized to fermionic superfluids~\cite{Zhang:2019}.
Unlike the mass current case, the Josephson current is expected not to contribute to the spin current.
Another generalization is the study beyond the linear response theory to discuss the good-contact regime.
While such an analysis is generically complicated, the quasiparticle current takes the form of Eq.~(\ref{eq:Ispin}) up to a prefactor in some situation (see Appendix~\ref{appendix:nonlinear}).
Our formalism also predicts the noise of the spin current.
Within the linear response theory, the noise  at zero frequency is related to $I_\mathrm{spin}$ by $S(\omega=0)=\coth(h/T)I_\mathrm{spin}$ as derived in Appendix~\ref{appendix:noise}.
We expect such a noise to be accessible in future ultracold atom experiments~\cite{Uchino:2018}.

\acknowledgments
The authors are supported by JSPS KAKENHI Grant Number JP17K14366, JP17J03975, JP18H05406, and JP19J01006.
S.U. is also supported by a Waseda University Grant for Special Research Projects (No. 2019C-461).

\appendix
\section{\label{appendix:noise}Linear response and noise}
This appendix is devoted to the linear response theory to the tunneling amplitude $t$.
For convenience, we introduce $C_\sigma\equiv t\sum_{\p,\p'}c_{\p,\sigma,R}^\+c_{\p',\sigma,L}$ and $\hat{I}_\sigma(t')\equiv-dN_{\sigma,L}(t')/dt'$.
The mass and spin current operators are given by $\hat{I}_\mathrm{mass}=\hat{I}_\up+\hat{I}_\down$ and $\hat{I}_\mathrm{spin}=\hat{I}_\up-\hat{I}_\down$, respectively.
The tunneling Hamiltonian in Eq.~(\ref{eq:H_T}) is rewritten as $H_T=\sum_{\sigma=\up,\down}(C_\sigma+C_\sigma^\+)$.
Using the Heisenberg equation combined with $\{c_{\p,\sigma,j},c^\+_{\p',\sigma',j'}\}=\delta_{\p\p'}\delta_{\sigma\sigma'}\delta_{j j'}$ and $\{c_{\p,\sigma,j},c_{\p',\sigma',j'}\}=0$, we obtain 
$\hat{I}_\sigma(t')= i[N_{\sigma,L},H]=-iC_\sigma+iC_\sigma^\+.$

First, we review the linear response of the current for the spin-$\sigma$ component.
We follow the procedure in Ref.~\cite{Mahan:2013}.
Hereafter, $\<\cdots\>$ denotes the expectation value for a given non-equilibrium state.
According to the Kubo formula, $I_\sigma\equiv\<\hat{I}_\sigma(t')\>$ for a steady state is given by
\begin{align}\label{eq:<I_up>}
I_\sigma&=-i\int_{-\infty}^{t'}\!\!\!\!\!dt''\,\<[\hat{I}_\sigma^{(H_0)}(t'),H_T^{(H_0)}(t'')]\>_\mathrm{eq}+O(t^3),
\end{align}
where $\<\cdots\>_\mathrm{eq}$ is a thermal average for the Fermi gases in both reservoirs, and $O^{(A)}(t')\equiv e^{iAt'}Oe^{-iAt'}$ has been defined.
Operators in the right hand side are in the Heisenberg representation for $H_0=K_L+K_R+\sum_{\sigma,j}\mu_{\sigma,j}N_{\sigma,j}$.
Using the Baker Campbell Hausdorff formula, we can find $C_\sigma^{(H_0)}(t')=e^{-i\Delta\mu_\sigma t'}C_\sigma^{(K_0)}(t')$, where $K_0=K_L+K_R$ and $\Delta\mu_\sigma=\mu_{\sigma L}-\mu_{\sigma R}$.
(Henceforth, the shorthand notation $C_\sigma(t')=C_\sigma^{(K_0)}(t')$ is used only for the operator $C_\sigma$.)
By employing this as well as the expressions of $\hat{I}_\sigma$ and $H_T$ in terms of $C_\sigma$, Eq.~(\ref{eq:<I_up>}) in the normal phase can be rewritten as
\begin{align}\label{eq:I_sigma}
I_\sigma&=2\mathrm{Im}\left[\chi_{C_\sigma}^\mathrm{ret}(-\Delta\mu_\sigma)\right],
\end{align}
where
\begin{align}\label{eq:chi^ret}
\chi_{C_{\sigma}}^\mathrm{ret}(\omega)&=-i\int_0^\infty\!dt'\,e^{i\omega t'}\<[C_\sigma(t'),C_{\sigma}^\+(0)]\>_\mathrm{eq}
\end{align}
is a retarded correlation function.
By using the commutability of $K_L$ and $K_R$ in Eq.~(\ref{eq:Hamiltonian}) as well as an interrelation between retarded and Matsubara Green's functions, $I_\mathrm{spin}$ can be rewritten as Eq.~(\ref{eq:Ispin}) in terms of DOSs.

Let us now turn to a noise in mesoscopic transport.
The noise is defined as a spectral density of current fluctuations.
In the case of the spin current, this is given by
\begin{align}\label{eq:def:S}
S(\omega)&=\int\!dt'\,e^{i\omega t'} \<\delta\hat{I}_\mathrm{spin}(t')\delta\hat{I}_\mathrm{spin}(0)\>,
\end{align}
where $\delta\hat{I}_\mathrm{spin}=\hat{I}_\mathrm{spin}-\<\hat{I}_\mathrm{spin}\>$.
Since $\<\hat{I}_\mathrm{spin}\>=O(t^2)$, the expectation value in Eq.~(\ref{eq:def:S}) to leading order is equivalent to an equilibrium current correlation.
In addition, cross correlation functions between currents with opposite spins vanish in the normal phase.
Therefore, Eq.~(\ref{eq:def:S}) reduces to
\begin{align}\label{eq:S_1}
S(\omega)&=\sum_\sigma\int\!dt'\,e^{i\omega t'}\<\hat{I}_\sigma^{(H_0)}(t')\hat{I}_\sigma^{(H_0)}(0)\>_\mathrm{eq}+O(t^3).
\end{align}
As shown above, we have $\hat{I}_\sigma=-iC_\sigma+iC_\sigma^\+$ and $C_\sigma^{(H_0)}(t')=e^{-i\Delta\mu_\sigma t'}C_\sigma(t')$.
Furthermore, the thermal averages of $C_{\sigma}(t')C_{\sigma}(0)$ and its Hermitian conjugate vanish in the normal phase.
As a result, Eq.~(\ref{eq:S_1}) reads
\begin{align}\label{eq:S_2}
S(\omega)&=\sum_\sigma\left[
\chi_{C_{\sigma}}^>(\omega-\Delta\mu_\sigma)
+\chi_{C_{\sigma}}^<(-\omega-\Delta\mu_\sigma)
\right],
\end{align}
where greater and lesser correlation functions are given by
\begin{align}\label{eq:chi^>}
\chi_{C_{\sigma}}^>(\omega')&=\int\!dt'\,e^{i\omega' t'} \<C_{\sigma}(t')C_{\sigma}^\+(0)\>_\mathrm{eq},\\
\label{eq:chi^<}
\chi_{C_{\sigma}}^<(\omega')&=\int\!dt'\,e^{i\omega' t'} \<C_{\sigma}^\+(0)C_{\sigma}(t')\>_\mathrm{eq},
\end{align}
respectively.
The Lehmann representations of Eqs.~(\ref{eq:chi^ret}), (\ref{eq:chi^>}), and (\ref{eq:chi^<}) provide the following relation:
\begin{align}
\chi_{C_{\sigma}}^>(\omega')
&=e^{\omega'/T}\chi_{C_{\sigma}}^<(\omega')
=\frac{2\mathrm{Im}\left[\chi_{C_\sigma}^\mathrm{ret}(\omega')\right]}{e^{-\omega'/T}-1}.
\end{align}
Combining this with Eqs.~(\ref{eq:I_sigma}) and (\ref{eq:S_2}), we obtain
\begin{align}\label{eq:S(0)}
S(0)&=\sum_\sigma\coth\left(\frac{\Delta\mu_\sigma}{2T}\right)I_\sigma.
\end{align}
In the case of the chemical potentials shown in Fig.~\ref{fig:setup}, we have $\Delta\mu_\up=-\Delta\mu_\down=2h$.
Therefore, the noise is related to $I_\mathrm{spin}$ by $S(0)=\coth(h/T)I_\mathrm{spin}$.

In the end of this appendix, we will comment on the noise of the mass current, which is given by replacing $\delta\hat{I}_\mathrm{spin}$ in Eq.~(\ref{eq:def:S}) with $\delta\hat{I}_\mathrm{mass}=\hat{I}_\mathrm{mass}-\<\hat{I}_\mathrm{mass}\>$.
As mentioned above, cross correlation functions between currents with opposite spins vanish in the normal phase.
As a result, both mass- and spin-current noises have the same form of Eq.~(\ref{eq:S_2}).
When a bias is spin independent ($\mu_{\sigma,j}=\mu_j,\,\Delta\mu=\mu_L-\mu_R\neq0$), there is no spin current and the noise is related to the mass current by $S(0)=\coth[\Delta\mu/(2T)]I_\mathrm{mass}$.

\section{\label{appendix:free}$I_\mathrm{spin}$ and $G_\mathrm{spin}$ for free fermions}
Here, the spin current [Eq.~(\ref{eq:Ispin})] in the absence of interparticle interactions is considered.
We assume $\mu_+=\mu+h>0$ and $h>0$.
In this case, the density of states in Eq.~(\ref{eq:DOS}) reduces to
\begin{align}
\label{freeDOS}
\rho_\pm(\omega)&=\frac{m}{2\pi^2}\sqrt{2m(\omega+\mu_\pm)}\theta(\omega+\mu_\pm),
\end{align}
where $\theta(\omega)$ is the Heaviside step function.
Substituting this into Eq.~(\ref{eq:Ispin}), we obtain
\begin{align}\label{eq:I_spin^(0)}
I_\mathrm{spin}
&=I_\mathrm{spin}^{0}\equiv
\frac{2m^3t^2T^2}{\pi^3}\sum_{\alpha=\pm}(-\alpha)\mathrm{Li}_2(-e^{\mu_\alpha/T}),
\end{align}
where $\mathrm{Li}_n(z)\equiv\sum_{k=1}^\infty\frac{z^k}{k^n}$ is the polylogarithm.
The zero-bias spin conductance in Eq.~(\ref{eq:G_spin}) reads
\begin{align}
G_\mathrm{spin}|_{h\to0}&=\frac{2m^3t^2T}{\pi^3}\ln(1 + e^{\mu/T}).
\end{align}
On the other hand, the large-$h$ behavior of $I_\mathrm{spin}^{0}$ with $\mu$ and $T$ fixed is
\begin{align}
I_\mathrm{spin}^{0}
&=\frac{m^3t^2T^2}{\pi^3}\left(\mu_+^2+\frac{\pi^2 T^2 }{3}+O(e^{-h/T})\right).
\end{align}
At $T=0$, we find 
$I_\mathrm{spin}^{0}=m^3t^2[\mu_+^2-\mu_-^2\theta(\mu_-)]/\pi^3$ and $G_\mathrm{spin}|_{h\to0}=G_\mathrm{spin}^{0}\equiv2m^3t^2\varepsilon_{\mathrm{F},0}/\pi^3$.
Note $\mu=\varepsilon_{\mathrm{F},0}$ at zero temperature in the non-interacting case.

\section{\label{appendix:nonlinear}Nonlinear response}
Here we discuss how non-equilibrium properties beyond the linear response affect the spin transport and show that our analysis in this work is not largely affected by them.
By collecting the single-particle tunneling process,
the non-linear quasi-particle spin current $I_{\rm spin}$ is obtained as~\cite{Uchino:2017}
\begin{eqnarray}
\label{eqC1}
I_{\rm spin}&=&4\pi t^2\int d\omega\frac{\rho_{+}(\omega-h)\rho_{-}(\omega+h)}{|1-4\pi^2t^2g_{+}(\omega-h)g_{-}(\omega+h)|^2}\cr
&&\times\left[f(\omega-h)-f(\omega+h)\right].
\end{eqnarray}
The definition of $g_\alpha(\omega)$ is given by
\begin{eqnarray}
g_\alpha(\omega)&=&\sum_{\bm{p}}^{|\bm{p}|<\Lambda}\mathcal{G}_\alpha(\bm{p},i\omega_n\rightarrow\omega+i\delta) ,
\end{eqnarray}
where $\Lambda$ is the momentum cutoff of the tunneling term.
The real part of $g_\alpha(\omega)$ involves the ultraviolet divergence.
When the momentum cutoff $\Lambda$ is enough large,
the self-energy corrections in the denominator of Eq.~(\ref{eqC1}) is irrelevant.
In the absence of the self-energy, we obtain 
\begin{eqnarray}
g_\alpha(\omega)=-\frac{m}{\pi^2}\left[\Lambda+i\frac{\pi}{2}\sqrt{2m(\omega+\mu_\alpha)}\right].
\end{eqnarray} 
We note that the real part is proportional to $\Lambda$ and the imaginary part gives the single-particle density of states. 
When the term proportional to the cutoff gives the dominant contribution, the denominator of the non-linear spin current equation~(\ref{eqC1}) gives
\begin{eqnarray}
|1-4\pi^2t^2g_{+}(\omega-h)g_{-}(\omega+h)|^2\simeq\left|1-\frac{4m^2t^2\Lambda^2}{\pi^2}\right|^2.
\end{eqnarray}
Therefore the non-linear spin current reads
\begin{eqnarray}
I_{\rm spin}&\simeq&\frac{4\pi t^2}{\left|1-\frac{4m^2t^2\Lambda^2}{\pi^2}\right|^2}\int d\omega\rho_{+}(\omega-h)\rho_{-}(\omega+h)\cr
&&\times\left[f(\omega-h)-f(\omega+h)\right],
\end{eqnarray}
which is the same form of the linear response except for the coefficient.
The current expression requires that $t$ is small such that $mt\Lambda\ll1$.
These parameters would be chosen to reproduce the detailed experimental setup of the quantum point contact.
If $\Lambda$ is comparable to the imaginary part, namely DOS,
we have to seriously consider the corrections from both real and imaginary parts of $g_\alpha(\omega)$.
It will be left as an important future work.


\begin{thebibliography}{99}
\bibitem{note}
We note that, although analytical forms of both the green dotted and blue dashed lines in Fig.~\ref{fig:Ispin@unitarity} are given by Eq.~\ref{eq:I_spin^(0)}, different $\mu$ are substituted.
In the case of the former, $\mu=0.94\varepsilon_{\mathrm{F},0}$ is the non-interacting chemical potential at $T=0.25T_{\mathrm{F},0}$, i.e., determined by solving Eq.~(\ref{eq:number}) with $(k_{\mathrm{F},0}a)^{-1}=-\infty$.
On the other hand, $\mu=0.33\varepsilon_{\mathrm{F},0}$ for the latter is obtained by solving Eq.~(\ref{eq:number}) at unitarity and is identical to that used in the red solid line in Fig.~\ref{fig:Ispin@unitarity}.
\end{thebibliography}
\end{document}